\documentclass[aps,prb,twocolumn,showpacs,preprintnumbers,superscriptaddress]{revtex4}
\usepackage{graphicx}
\usepackage{latexsym}
\usepackage{amssymb}
\begin{document}
\date{\today}

\title{Strong Pauli-limiting behavior of $H_{c2}$ and uniaxial pressure dependencies in KFe$_{\mathrm{2}}$As$_{\mathrm{2}}$}

\author{P. Burger}
\affiliation{Institut f\"ur Festk\"orperphysik, Karlsruher Institut f\"ur Technologie, 76021 Karlsruhe, Germany}
\affiliation{Fakult\"at f\"ur Physik, Karlsruher Institut f\"ur Technologie,  76131 Karlsruhe, Germany}

\author{F. Hardy} 
\affiliation{Institut f\"ur Festk\"orperphysik, Karlsruher Institut f\"ur Technologie, 76021 Karlsruhe, Germany}

\author{D. Aoki}
\affiliation{INAC, SPSMS, CEA Grenoble, 38054 Grenoble, France}
\affiliation{IMR, Tohoku University, Oarai, Ibaraki 311-1313, Japan}

\author{A. E. B\"ohmer}
\affiliation{Institut f\"ur Festk\"orperphysik, Karlsruher Institut f\"ur Technologie, 76021 Karlsruhe, Germany}
\affiliation{Fakult\"at f\"ur Physik, Karlsruher Institut f\"ur Technologie,  76131 Karlsruhe, Germany}

\author{R. Eder} 
\affiliation{Institut f\"ur Festk\"orperphysik, Karlsruher Institut f\"ur Technologie, 76021 Karlsruhe, Germany}

\author{R. Heid} 
\affiliation{Institut f\"ur Festk\"orperphysik, Karlsruher Institut f\"ur Technologie, 76021 Karlsruhe, Germany}

\author{T. Wolf}
\affiliation{Institut f\"ur Festk\"orperphysik, Karlsruher Institut f\"ur Technologie, 76021 Karlsruhe, Germany}

\author{P. Schweiss}
\affiliation{Institut f\"ur Festk\"orperphysik, Karlsruher Institut f\"ur Technologie, 76021 Karlsruhe, Germany}

\author{R. Fromknecht}
\affiliation{Institut f\"ur Festk\"orperphysik, Karlsruher Institut f\"ur Technologie, 76021 Karlsruhe, Germany}

\author{M. J. Jackson}
\affiliation{Institut N\'{e}el, CNRS and Universit\'{e} Joseph Fourier, BP 166, 38042 Grenoble Cedex 9, France}

\author{C. Paulsen}
\affiliation{Institut N\'{e}el, CNRS and Universit\'{e} Joseph Fourier, BP 166, 38042 Grenoble Cedex 9, France}

\author{C. Meingast}
\affiliation{Institut f\"ur Festk\"orperphysik, Karlsruher Institut f\"ur Technologie, 76021 Karlsruhe, Germany}

\begin{abstract}
KFe$_{2}$As$_{2}$ single crystals are studied using specific-heat, high-resolution thermal-expansion, magnetization, and magnetostriction measurements. The magnetization and magnetostriction data provide clear evidence for strong Pauli limiting effects of the upper critical field for magnetic fields parallel to the FeAs planes, suggesting that KFe$_{2}$As$_{2}$ may be a good candidate to search for the Fulde-Ferrell-Larkin-Ovchinnikov (FFLO) state. Using standard thermodynamic relations, the uniaxial pressure derivatives of the critical temperature ($T_{c}$), the normal-state Sommerfeld coefficient ($\gamma_{n}$), the normal-state susceptibility ($\chi$), and the thermodynamic critical field ($H_{c}$) are calculated from our data. We find that the close relationship between doping and pressure as found in other Fe-based systems does not hold for KFe$_{2}$As$_{2}$.
\end{abstract}


\maketitle

\section{Introduction}\label{sec:intro}
The detailed understanding of Fe-based superconductors continues to present a considerable challenge in condensed matter physics.~\cite{Johnston10,Paglione10,Stewart11,Mazin09} Of particular interest recently has been the strongly hole doped compound KFe$_{2}$As$_{2}$, which is the end member of the (Ba,K)Fe$_{2}$As$_{2}$ system and has a much lower $T_{c}$ of only 3.4 K than the optimal $T_{c}$ value of about 40 K near 40\%  K content.~\cite{Fukazawa09,Terashima09,Rotter08} Whereas Ba$_{0.6}$K$_{0.4}$Fe$_{2}$As$_{2}$ appears to have a fully gapped $s$-wave order parameter,~\cite{Popovich10,Li11} there are indications for a nodal superconducting state in KFe$_{2}$As$_{2}$. In fact, a $d$-wave state was predicted early on from functional renormalization group theory.~\cite{Thomale11} Experimentally, penetration depth and thermal conductivity studies have been interpreted in terms of a $d$-wave order parameter,~\cite{Hashimoto10,Reid12} whereas recent laser angle-resolved photoemission (ARPES) experiments suggest a nodal $s$-wave state.~\cite{Okazaki12} The electronic structure of KFe$_{2}$As$_{2}$ has been investigated both with de Haas-van Alphen (dHvA) and ARPES methods,~\cite{Terashima09,Sato09} and these studies show that KFe$_{2}$As$_{2}$ has only hole pockets.  This of course immediately raises the question of whether superconductivity in KFe$_{2}$As$_{2}$ has the same mechanism as the optimally doped system, for which superconductivity has been suggested to originate from interband pairing between electron and hole pockets.~\cite{Mazin09} Inelastic neutron scattering results still show signs of spin fluctuations, which are however incommensurate but may nevertheless lead to superconducting pairing in the heavily overdoped region.~\cite{Lee11,Castellan11} Paradoxically, KFe$_{2}$As$_{2}$ has the largest $\gamma_{n}$ in the (Ba,K)Fe$_{2}$As$_{2}$ system in spite of the low $T_{c}$ value, and this has been linked to a close proximity to an orbitally-selective Mott transition due to strong Hund correlations.~\cite{Hardy13,deMedici12}\\  

In this Article we study the normal- and superconducting-state properties of KFe$_{2}$As$_{2}$ using several thermodynamic probes: specific-heat, high-resolution thermal-expansion, magnetization, and magnetostriction measurements. Our magnetization and magnetostriction data clearly show that KFe$_{2}$As$_{2}$ is strongly Pauli limited for fields parallel to the FeAs planes and, thus, this system may be another possible candidate to search for the Fulde-Ferrell-Larkin-Ovchinnikov~\cite{Fulde64,Larkin65} (FFLO) state. Using standard thermodynamic relations, the uniaxial pressure derivatives of the critical temperature ($T_{c}$), the normal-state Sommerfeld coefficient ($\gamma_{n}$), the normal-state magnetic susceptibility ($\chi$), and the thermodynamic critical field ($H_{c}$) are calculated from our data. First, we find that the uniaxial pressure derivatives of $T_{c}$ are very anisotropic and of opposite sign as compared to Co-doped Ba122.~\cite{Hardy09} We find that both $T_{c}$ and $\gamma_{n}$ decrease under hydrostatic pressure.  This is in contrast to the doping induced behavior and shows that pressure and doping can not be equated, as they can in the Co- and P-doped Ba122 systems.~\cite{Meingast12,Boehmer12}

\section{Experimental details}\label{sec:experimental}
Single crystals of KFe$_{2}$As$_{2}$ were grown in alumina crucibles using a K-As rich flux with a K:Fe:As ratio of about 0.3:0.1:0.6. The crucibles were sealed in an iron cylinder filled with argon gas. After heating up to 980$\,^{\circ}\mathrm{C}$ the furnace was cooled down slowly at a rate of about 0.5$\,^{\circ}\mathrm{C}$/h. Crystals with dimensions up to 3.0 $\times$ 2.5 $\times$ 1.0 mm$^{3}$ were used in the present investigation. The specific heat was measured with a commercial Quantum Design Physical Property Measurement System (PPMS) for $T$ $>$ 0.4 K and with a home-made calorimeter for $T$ $<$ 0.4 K. For $T$ $>$ 2 K, we used a vibrating sample magnetometer to measure the magnetization. For $T$ $<$ 2 K, magnetization measurements were performed using a low-temperature superconducting quantum interference device (SQUID magnetometer) equipped with a miniature dilution refrigerator developed at the Institut N\'eel-CNRS Grenoble. The sample was attached to a copper tress suspended from the dilution unit's mixing chamber which descends through the bore of the magnet. The magnetometer is equipped with a solenoid capable of producing fields up to 8 T. The setup can measure absolute values of the magnetization by the extraction method at temperatures down to 75 mK. The thermal expansion and the magnetostriction were measured in a 
custom-made capacitive dilatometer with a typical resolution of $\Delta L/L_{0}$ $\sim$ 10$^{-8}$-10$^{-10}$.~\cite{Meingast90,Meingast12}

\section{Results}\label{sec:results}
\subsection{Heat capacity}\label{sec:heatcapa}

\begin{figure}[h]
\begin{center}
\includegraphics[width=86mm]{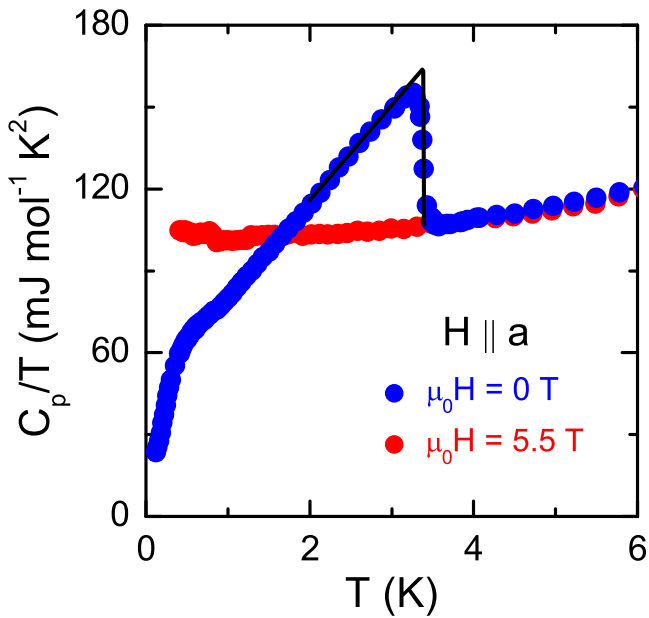}
\end{center}
\caption{(Color Online) Specific heat of KFe$_{2}$As$_{2}$ in 0 and 5.5 T for H $||$ $a$.}
\label{Fig1}
\end{figure}

Figure \ref{Fig1} shows typical heat-capacity data of our samples, clearly demonstrating a sharp superconducting transition at T$_{c}$ = 3.4 K and a large Sommerfeld coefficient $\gamma$ $\approx$ 100 mJ mol$^{-1}$ K$^{-2}$, as reported earlier.~\cite{Hardy13} We attribute the decrease of $C_{p}/T$ below about 0.5 K to small superconducting gaps on parts of the Fermi surface, which surprisingly appear to be absent in recent data.~\cite{Grinenko13} In the following, we use the data of Fig. \ref{Fig1} for calculating the thermodynamical critical field (see Section ~\ref{sec:Hc}) and the uniaxial pressure dependences (see Section ~\ref{sec:discussion2}).

\subsection{Reversible magnetization}\label{sec:magnet}
\begin{figure}[t]
\begin{center}
\includegraphics[width=83mm]{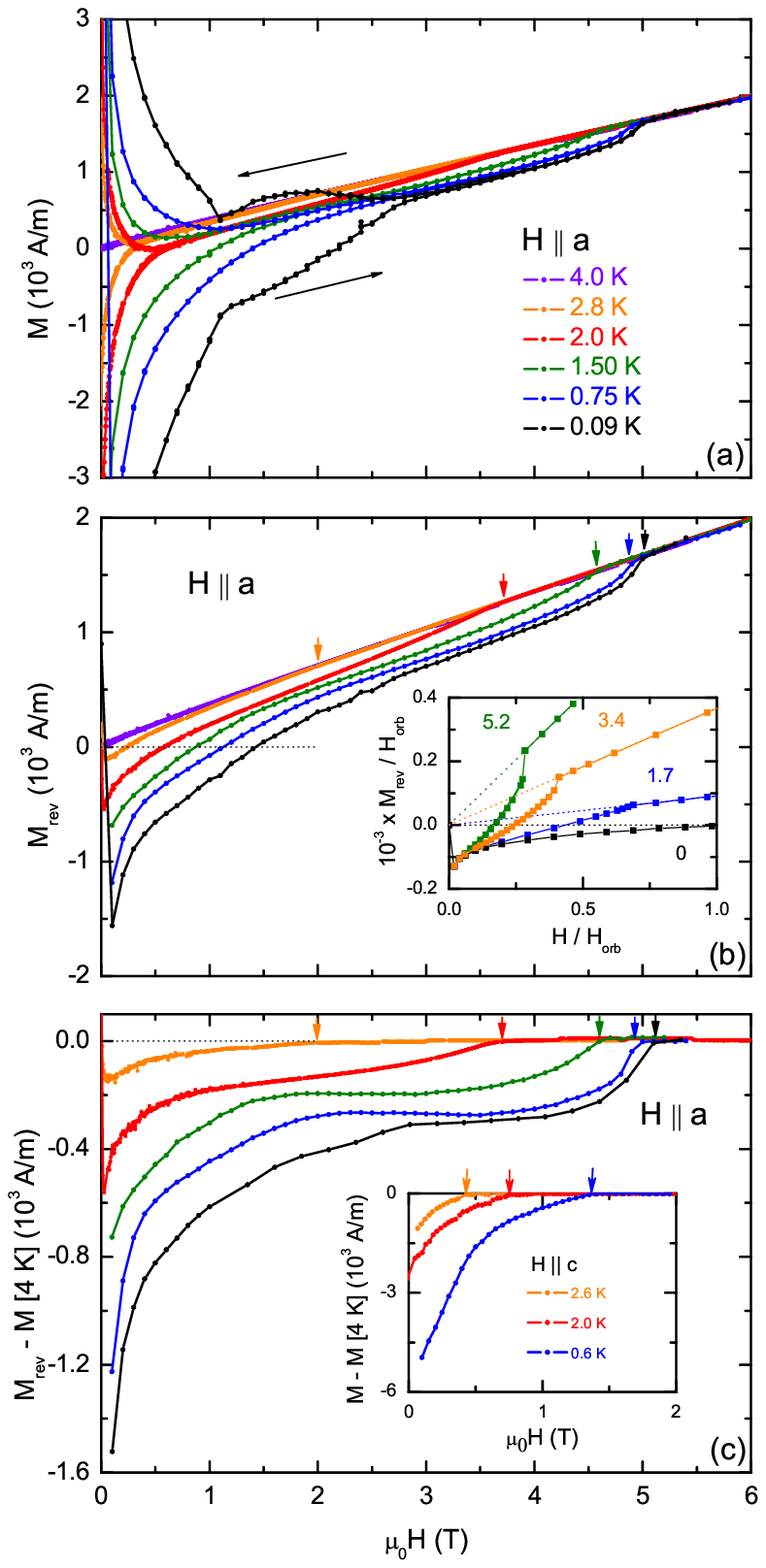}
\end{center}
\caption{(Color Online) (a) Magnetization curves for H $||$ $a$ at different temperatures. (b) Reversible part of the magnetization. The inset shows calculations from Ichioka and Machida~\cite{IchiokaBook,Ichioka07} for several values of $\alpha_{M}$ (for $\alpha_{M}$ $>$ 1.8, transitions are 1$^{st}$ order). (c) Difference between the reversible superconducting- and normal-state magnetizations. The inset shows data for H $||$ c at several temperatures. The arrows indicate the values of H$_{c2}$(T).}
\label{Fig2}
\end{figure}

Figure \ref{Fig2}(a) shows raw magnetization curves measured down to 0.09 K in increasing and decreasing magnetic field applied parallel to the $a$ axis. In the normal state, {\it i.e.} for T = 4 K, we find a large paramagnetic contribution with a field-independent susceptibility $\chi_{a}$ of about 4 $\times$ 10$^{-4}$ in agreement with our previous study.~\cite{Hardy13} At all temperatures, the magnetization is fully reversible over a wide field interval below the upper critical field H$_{c2}$(T). This shows that our samples have very weak flux pinning, which is compatible with the recent observation of a well-defined hexagonal vortex lattice,~\cite{Kawano11} but is in strong contrast to more disordered Co-doped systems.~\cite{Eskildsen09,Yamamoto09} Thus, accurate reversible magnetization data can be obtained by averaging the field increasing and decreasing branches of the magnetization loop, as shown in Fig.\ref{Fig2}(b). These curves clearly exhibit several features characteristic of strongly Pauli-limited superconductors.~\cite{Tayama02,Paulsen11,Tenya06}
First, M(H) is negative, {\it i.e.} diamagnetic, only in a narrow low field interval 0 $<$ H/H$_{c2}$ $<$ 0.3 and positive, {\it i.e.} paramagnetic, for higher fields. Second, for the lowest temperature measured (T = 0.09 K), M(H) increases strongly upon approaching H$_{c2}$(T), rather than exhibiting the linear behavior expected from Ginzburg-Landau theory in the abscence of paramagnetic effects. This is more clearly seen in Fig.\ref{Fig2}(c), where the normal-state magnetization has been subtracted. Similar behavior was already reported in both high-$\kappa$ dirty Ti-V alloys ($\kappa$ $\approx$ 68)~\cite{Hake67} and clean CeCoIn$_{5}$ ($\kappa$ $\approx$ 100)~\cite{Tayama02,Paulsen11} and represent direct evidence for the existence of strong paramagnetic effects in KFe$_{2}$As$_{2}$ which  become more and more important with decreasing temperatures.
 
The behavior of Pauli-limited superconductors was investigated theoretically in detail by Ichioka and Machida.~\cite{IchiokaBook,Ichioka07} In the inset of Fig.\ref{Fig2}(b), we show their calculations (at T/T$_{c}$ = 0.1) for different values of the Maki parameter,
\begin{equation}
\alpha_{M}=\sqrt{2}\frac{H_{orb}(0)}{H_{p}(0)},
\label{eq1}
\end{equation} 
(where H$_{orb}$(0) and H$_{p}$(0) are the zero-temperature values of the orbital and Pauli fields, respectively) which is a measure of the paramagnetic pair-breaking strength.~\cite{Maki64} These theoretical curves clearly show that {\it (i)} the magnetization becomes rapidly positive with increasing field in the superconducting state with increasing $\alpha_{M}$ and {\it (ii)} the transition to the normal state at H$_{c2}$ becomes first-order for large values of $\alpha_{M}$. Our 0.09 K curve (T/T$_{c}$ $\approx$ 0.03) lies somewhere between the calculated curves for $\alpha_{M}$ = 1.7 and 3.4 suggesting a weakly first-order transition at T = 0 K. Thus, our measurements clearly show, for the first time, the existence of strong paramagnetic depairing effects in KFe$_{2}$As$_{2}$ for H $||$ $a$. As shown in the inset of Fig.\ref{Fig2}(c), there is almost no paramagnetic effect for H $||$ $c$ and $\alpha_{M}$ $\approx$ 0 in this direction.  

\subsection{Thermodynamic critical field}\label{sec:Hc}                     
Hereafter, we show that the thermodynamic critical field $H_{c}$(T) obtained from the heat capacity data matches that obtained by our reversible magnetization measurements quite well. $H_{c}$(T), which measures the Cooper-pairs condensation energy, can be determined directly from zero-field heat-capacity data using
\begin{equation}
-\frac{\mu_{0}}{2} H^{2}_{c}(T) = \int^{T}_{0} (S_{s}(T') - S_{n}(T')) dT',
	\label{eq2}
\end{equation}
(where $S_{n}$ and $S_{s}$ are the normal- and superconducting-state entropies, respectively) or using reversible magnetization curves with
\begin{equation}
-\frac{\mu_{0}}{2}H^{2}_{c}(T)=\mu_{0}\int^{H_{c2}}_{0} (M_{s}(T) - M_{n}(T)) dH , 
	\label{eq3}
\end{equation}
where $M_{n}$ and $M_{s}$ are the normal- and superconducting-state magnetizations, respectively.
\begin{figure}[b]
\begin{center}
\includegraphics[width=86mm]{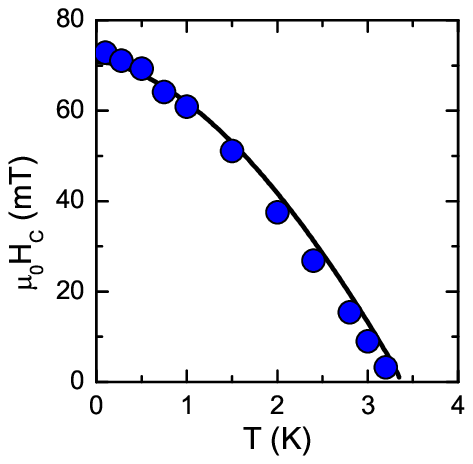}
\end{center}
\caption{(Color Online) Temperature dependence of the thermodynamic critical field inferred from specific-heat (line) and magnetization (symbol) measurements.}
\label{Fig3}
\end{figure}
The resulting values of $H_{c}$(T) (see Fig.\ref{Fig3}) inferred from magnetization data agree quite well with those calculated from the heat capacity, demonstrating the overall consistency between our thermodynamic measurements. It also indicates that the heat capacity is not contaminated by any spurious disordered magnetic contributions, as reported in Refs~\onlinecite{Hafiez12,Grinenko13}. In Section \ref{sec:discussion}, we use the derived $H_{c}$(T) to discuss the (H,T) phase diagram of KFe$_{2}$As$_{2}$. 

\subsection{Thermal expansion and magnetostriction}\label{sec:alpha}
\begin{figure}[h]
\begin{center}
\includegraphics[width=86mm]{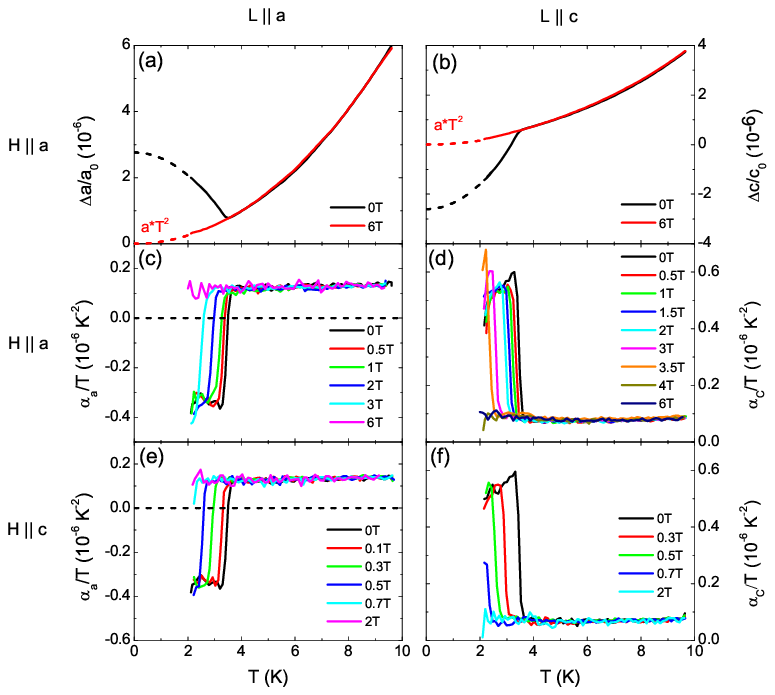}
\end{center}
\caption{ (Color Online) (a) Relative length change versus temperature of the $a$-axis (b) and $c$-axis in the normal (red curves) and the superconducting state (black curves). (c) - (f) Uniaxial thermal-expansion coefficients divided by temperature $\alpha_{a, c}/T$ as a function of temperature for both crystallographic axes and different fields.}
\label{Fig4}
\end{figure}

We also performed thermal-expansion and magnetostriction measurements in order to study the effects of pressure, in particular uniaxial pressure, on the superconducting- and normal-state properties of KFe$_{2}$As$_{2}$. Clear anomalies in the relative length changes $\Delta L_{i}/L_{0}$ ($i$=$a,c$) are seen at $T_{c}$($H$ = 0 T) = 3.4 K as shown in Fig.\ref{Fig4}(a) and (b). The red curves indicate the normal-state behavior, where superconductivity has been suppressed by a field of $H$ = 6 T applied along the $a$-axis, and the dashed lines indicate the extrapolated behavior down to $T$ = 0 K.

The structural distortions in the superconducting state provide a direct indication of how the system can lower its free energy, and from these data it is clear that superconductivity favors a longer (shorter) $a$-axis ($c$-axis). Fig.\ref{Fig4}(c) - (f) show the corresponding uniaxial thermal-expansion coefficients $\alpha_{i} = (1/L_{i}) dL_{i}/dT$ ($i = a, c$), divided by temperature at different applied magnetic fields along the $a$-axis ((c) and (d)) and the $c$-axis ((e) and (f)). The anomalies at $T_{c}$ have a clear step-like shape, indicating second-order phase transitions and decrease in temperature with increasing magnetic field.  

Above T$_{c}$, $\alpha_{i}/T$ is constant and field independent, as expected for a Fermi liquid. Using Maxwell relations this value is related to the pressure dependence of $\gamma_{n}$:~\cite{Meingast12}
\begin{equation}
	\frac{\alpha_{i}}{T}=-\frac{1}{V}\frac{\partial (\frac{S}{T})}{\partial p_{i}}=-\frac{1}{V_{m}}\frac{d\gamma_{n}}{d p_{i}}, 
	\label{eq4}
\end{equation}
where $i$=$a,c$. The normal-state values of $\alpha_{i}/T$ are $\alpha_{a}/T = 0.12 \times 10^{-6}$ K$^{-2}$ and $\alpha_{c}/T = 0.08 \times 10^{-6}$ K$^{-2}$ which yield the following uniaxial-pressure dependencies of the normal-state Sommerfeld coefficients $d\gamma_{n}/dp_{a} = -7.74$ mJ mol$^{-1}$ K$^{-2}$ GPa$^{-1}$ and $d\gamma_{n}/dp_{c} = -4.81$ mJ mol$^{-1}$ K$^{-2}$ GPa$^{-1}$.

In order to determine the uniaxial-pressure dependencies of $T_{c}$, we use the Ehrenfest relation:~\cite{Ehrenfest38,Hardy09}
\begin{equation}
	\frac{dT_{c}}{dp_{i}}=\frac{\Delta \alpha_{i}V_{m}}{\Delta C_{p}/T_{c}},
	\label{eq5}
\end{equation}
where $i$=$a,c$. Here $\Delta \alpha_{i}$ is the jump in the thermal expansion along the $i$ direction, $V_{m} = 61.27$ cm$^{3}$ mol$^{-1}$ is the molar volume and $\Delta C_{p}$ is the specific heat jump.
Using our values for the thermal-expansion ($\Delta \alpha_{a} = -1.68$ $\times$ 10$^{-6}$ K$^{-1}$ and  $\Delta \alpha_{c} = 1.85$ $\times$ 10$^{-6}$ K$^{-1}$) and heat-capacity ($\Delta C_{p}/T_{c}$ = 54 mJ mol$^{-1}$ K$^{-2}$) jumps from Fig.\ref{Fig1}, we find $dT_{c}/dp_{a} = -1.92$ K GPa$^{-1}$ and $dT_{c}/dp_{c} = 2.10$ K GPa$^{-1}$. Our results show that the pressure dependence of $T_{c}$ in KFe$_{\mathrm{2}}$As$_{\mathrm{2}}$ is very anisotropic.  Indeed it is negative along the $a$-axis and positive along the  $c$-axis, although the magnitudes are comparable. Interestingly, superconductivity couples strongly to the $c/a$ ratio as in Co- and P-doped Ba122,~\cite{Hardy09,Boehmer12} but with opposite sign, {\it i.e.} a smaller, rather than a larger, $c/a$ ratio enhances $T_{c}$. Under hydrostatic conditions we get a negative $dT_{c}/dp_{vol} = 2 dT_{c}/dp_{a} + dT_{c}/dp_{c} = -1.74$ K GPa$^{-1}$. Our results are in qualitative agreement with the recent data of Bud'ko {\it et al.},~\cite{Budko12} who find $dT_{c}/dp_{vol} = -1.0$ K GPa$^{-1}$, $dT_{c}/dp_{a}$ $\approx -1.1$ K GPa$^{-1}$ and $dT_{c}/dp_{c}$ $\approx 1.1$ K GPa$^{-1}$ from hydrostatic pressure and $c$-axis thermal expansion data.


The pressure dependence of the thermodynamic critical field can be calculated using the following relation:~\cite{Shoenberg1962}
\begin{equation}
	\frac{\Delta L_{i}}{L_{i}} = \frac{L_{n, i}-L_{s, i}}{L_{s, i}} = \mu_{0} H_{c} \left(\frac{d H_{c}}{d p_{i}}\right), 
	\label{eq6}
\end{equation}
where $i$= $a$, $c$ and $\Delta L_{i}/L_{i}$ are the relative length changes from Fig.\ref{Fig4}. We obtain $dH^{a}_{c}/dp_{a} = -0.049$ T GPa$^{-1}$ and $dH^{a}_{c}/dp_{c} = 0.046$ T GPa$^{-1}$.

Additional information about how uniaxial pressure affects both the normal- and superconducting-state properties of KFe$_{2}$As$_{2}$ can be obtained from magnetostriction measurements. The magnetostriction coefficients $\lambda_{i}$ are directly related to the uniaxial pressure dependences of the magnetization via
\begin{equation}
	\lambda_{i} = \frac{1}{L_{i}} \frac{dL_{i}}{dH} = - \frac{dM}{dp_{i}},
	\label{eq7}
\end{equation}
where $i$= $a$, $c$.
\begin{figure}[htbp]
\begin{center}
\includegraphics[width=86mm]{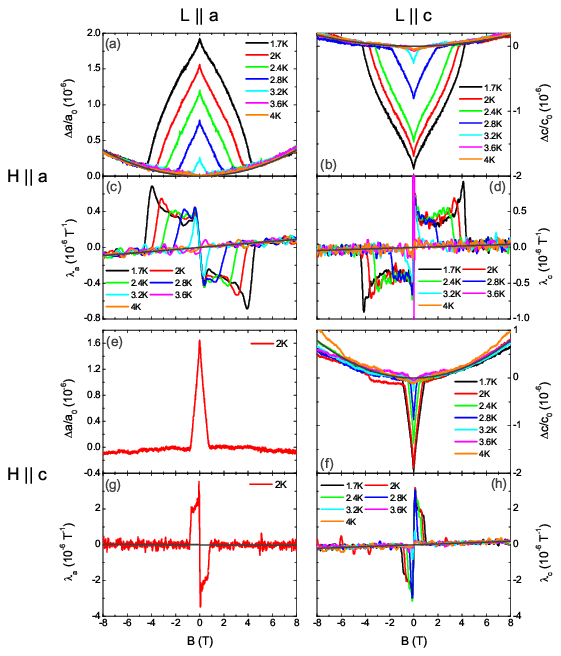}
\end{center}
\caption{ (Color Online) Magnetostriction $\Delta L_{i}(H)/L_{0}$ ($i$ = $a,c$) and linear magnetostriction coefficients $\lambda_{i}$(H) for the $a$- (left side) and $c$-axis (right side) for different field orientations. For H $||$ $a$, the anomaly in $\lambda_{i}$(H) at H$_{c2}$ increases strongly in magnitude with decreasing temperature (see (c) and (d)) due to the onset of Pauli-limiting behavior, whereas for H $||$ $c$ (see (g) and (h)), such an increase is notably absent.}
\label{Fig5}
\end{figure}
Figure \ref{Fig5} shows the magnetostriction data for various samples and field orientations for temperatures between 1.7 K and 4 K. The pressure dependence of the normal-state Pauli susceptibility can be directly obtained from $\lambda(H)$ for H$>$H$_{c2}$, which also varies linearly with field strength (see Fig.\ref{Fig5}(c), (d), (g), (h)) from Eq.\ref{eq7}. With this we extract $d\chi^{a}/dp_{a} = -1.43$ $\times$ 10$^{-5}$ GPa$^{-1}$, $d\chi^{c}/dp_{a} = 4.13$ $\times$ 10$^{-6}$ GPa$^{-1}$, $d\chi^{a}/dp_{c} = -6.74$ $\times$ 10$^{-6}$ GPa$^{-1}$, and $d\chi^{c}/dp_{c} = -3.08$ $\times$ 10$^{-5}$ GPa$^{-1}$. Interestingly, the biggest effects of uniaxial pressure on the magnetic susceptibility occur when the applied magnetic field is parallel to the pressure direction and is nearly one order of magnitude weaker for different orientations.

The magnetostriction below $T_{c}$ is fully reversible and thus, also provides information about how the reversible magnetization responds to uniaxial pressure.~\cite{Braendli68,Popovych06} The reversible magnetostriction depends basically on two parameters, $H_{c}$ and the Ginzburg-Landau parameter $\kappa$,  both of which can be pressure dependent. From the shape of $\lambda(H)$, it is evident that the main contribution to the magnetostriction comes from $dH_{c}/dp_{i}$ and not from $d\kappa/dp_{i}$ for both field directions.~\cite{Popovych06} For example, if $d\kappa/dp_{i}$ were the primary pressure derivative, $\lambda_{i}(H)$ would change sign near $H_{c2}/2$,~\cite{Popovych06} which is clearly not the behavior found in our data. For Pauli-limited superconductors, the Maki parameter may also be pressure dependent, complicating this simple analysis at high fields.  Indeed, for $H \parallel a$, we observe a strong increase in the size of the magnetostriction anomaly with decreasing temperatures, which is not expected for conventional superconductors.~\cite{Braendli68,Popovych06}  Since $\lambda$(H) is directly proportional to the pressure derivative of M(H) (see Eq.\ref{eq7}), this increase in the $\lambda$(H) anomaly directly reflects of the changing shape of the M(H) curve at low temperatures and signals the  crossover to a strongly Pauli-limited superconductor as the temperature is lowered.~\cite{Zocco} We also find an anomalous increase in the size of the expansivity anomaly for fields above about 3 T (see Fig.\ref{Fig4}(d)) which is also directly related to the onset of strong paramagnetic depairing for $H$ $||$ $a$. 

\section{Discussion}\label{sec:discussion}
\subsection{(H,T) phase diagram and paramagnetic effects}\label{sec:discussion1}
\begin{figure}[htbp]
\begin{center}
\includegraphics[width=86mm]{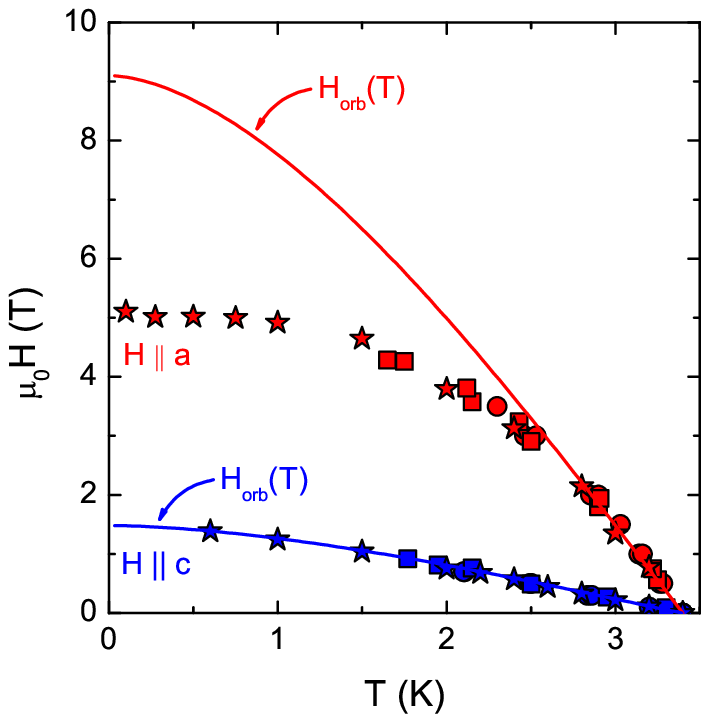}
\end{center}
\caption{ (Color Online) (H,T) phase diagram of KFe$_{2}$As$_{2}$ derived from magnetization ($\star$), thermal-expansion ($\circ$) and magnetostriction measurements ($\square$). Solid lines are the orbital fields calculated using the clean-limit Helfand-Werthamer theory. A strong suppression of $H_{c2}$ due to Pauli depairing is observed for H $||$ $a$.}
\label{Fig6}
\end{figure}

Figure \ref{Fig6} shows the superconducting (H,T) phase diagram of KFe$_{2}$As$_{2}$ derived from our thermodynamic measurements, which is similar to other results.~\cite{Terashima09,Liu13,Zocco} As expected, H$_{c2}$(T) is linear in the vicinity of T$_{c}$ for both field orientations, since the suppression of superconductivity in this region is always governed by the orbital effect. However, with decreasing temperature, it clearly flattens for H $||$ $a$, which corroborates the importance of paramagnetic depairing in this layered compound, already inferred from our magnetization curves. We note that a similar behavior is observed for nearly optimally K doped samples.~\cite{Altarawneh08} Although our heat-capacity data exhibit clear signs of multiband superconductivity (see Fig.\ref{Fig1}), we find no evidence of a sizeable change of curvature at high temperature due to the existence of several energy gaps, as reported in MgB$_{2}$ for H $\perp$ $c$.~\cite{Lyard04} The initial slopes $\left(\partial H_{c2}/\partial T\right)_{T_{c}}$ are equal to -0.6 and -3.7 T K$^{-1}$ for H $||$ $c$ and H $||$ $a$, respectively in agreement with the values of Terashima {\it et al.}~\cite{Terashima09} which lead to the coherence lengths $\xi_{GL}^{ab}$ $\approx$ 13 nm and $\xi_{GL}^{c}$ $\approx$ 2 nm. Using these values, we calculate the temperature dependence of the orbital field H$_{orb}$(T) for both directions using the Helfand-Werthamer theory~\cite{WHH1,WHH2,Brison95} in the clean limit. As shown in Fig.\ref{Fig6}, H$_{c2}^{c}$(T) is fully orbitally limited with H$_{c2}^{c}$(0) = H$_{orb}^{c}$(0) $\approx$ 1.5 T. For H $||$ $a$, Zeeman effects start to become significant below about 2.8 K and at T = 0 K, we find H$_{orb}^{ab}$(0) $\approx$ 9 T {\it i.e.} significantly larger than the measured H$_{c2}$(0) $\approx$ 5 T. Using H$_{c}$(0) = 73 mT (see Fig.\ref{Fig3}), we estimate the Maki-Ginzburg-Landau parameter $\kappa_{1}$ = $\frac{H_{orb}(0)}{\sqrt{2}\cdot H_{c}(0)}$ equal to 87 and 15 for H $||$ $a$ and H $||$ $c$, respectively. This shows that KFe$_{2}$As$_{2}$ is a very strong type II superconductor especially for H $||$ $a$. This is expected since paramagnetic effects are significant only in strong type II superconductors, in which the magnetic field strongly penetrates the sample.
The Pauli field H$_{p}$(0), {\it i.e.} the field at which the difference of Zeeman energy between the normal- and superconducting states exactly compensates the Cooper-pairs condensation energy, can be written in the following way~\cite{SaintJamesBook}
\begin{equation}
	H_{p}(0) = \frac{H_{c}(0)}{\sqrt{\chi_{n}-\chi_{s}}},
\label{eq8}
\end{equation}
Here $\chi_{n}$ and $\chi_{s}$ are the spin susceptibilities in the normal- and superconducting states, respectively. In the single-band $s$-wave case, $\chi_{s}$ = 0 and H$_{p}$(0) = 3.6 T (3.9 T) are obtained for H $||$ $a$  (H $||$ $c$). Here, we have used the values of $\chi_{n}$ obtained in Ref.~\onlinecite{Hardy13} (assuming implicitly that $\chi_{n}$ is dominated by the Pauli paramagnetism) and our value of H$_{c}$(0).  The above estimation, which is smaller than the measured H$_{c2}$(0) for H $||$ $a$, only provides a lower limit of the Pauli field, and H$_{p}$(0) can be enhanced by {\it e.g.} strong-coupling effects,~\cite{Orlando79} nodal gaps,~\cite{Yang98} or multiband superconductivity~\cite{Barzykin09}. In KFe$_{2}$As$_{2}$, an enhancement of H$_{p}$(0) by strong coupling can be discarded since $\Delta C_{p}/\gamma_{n}T_{c}$ $\approx$ 0.54 is substantially smaller than the single-band BCS value 1.43, in contrast to Ba$_{0.68}$K$_{0.32}$Fe$_{2}$As$_{2}$ where $\Delta C_{p}/C_{p}$ $\approx$ 2.5.~\cite{Popovich10} The response of a nodal superconductor to a magnetic field is quite different to that of a $s$-wave because the unpaired electrons located in the regions where the Zeeman field exceeds the local gap $\Delta(\bf{k})$ can be spin-polarized.~\cite{Yang98} Thus, $\chi_{s}$ is not zero and superconductivity can sustain a higher paramagnetic field. However, in this case $\chi_{s}$ $\approx$ 0.19$\cdot$$\chi_{n}$,~\cite{Won99} (neglecting the small field dependence of the gap) and H$_{p}$(0) is only slightly enhanced to 4 T for H $||$ $a$, which is still less than the observed H$_{c2}$(0).
\begin{table}[ht]
\caption{Normal- and superconducting-state parameters of KFe$_{2}$As$_{2}$} 
\centering 
\begin{tabular}{l c c} 
\hline\hline 
 & H $||$ a& H $||$ c \\  
\hline 
$\gamma_{n}$ (mJ mol$^{-1}$ K$^{-2}$)						& \multicolumn{2}{c}{103} 																		\\ 
T$_{c}$ (K)																			& \multicolumn{2}{c}{3.4}																			\\ 
$\chi_{n}$ 																			& 4.1$\times$10$^{-4}$ 						& 3.2$\times$10$^{-4}$ 			\\
$\kappa_{1}$ 																		&87																&15													\\
H$_{c}$(0) (T)& \multicolumn{2}{c}{0.073} 																																		\\
$\left(\partial H_{c2}/\partial T\right)_{T_{c}}$ (T K$^{-1}$)	&-3.7 &-0.6 																	\\
H$_{c2}$(0) (T) 																&5 																&1.5												\\
H$_{orb}$(0) (T) 																&9 																&1.5												\\
H$_{p}$(0) (T) 																	&\multicolumn{2}{c}{6.7}												\\
$\alpha_{M}$																		&1.8 															&0.3  												\\
\hline 
\end{tabular}
\label{table:resume} 
\end{table}
A significantly larger enhancement can be obtained in multiband superconductors with widely different gap amplitudes, i.e. $\Delta_{2}$ $>>$ $\Delta_{1}$.~\cite{Barzykin07,Barzykin09} In this scenario, the band with the smaller gap will almost recover its normal-state density of states, N$_{1}$(0), for H $>$ $\Delta_{1}$(0), while the second band, with the larger gap, will remain gapped all the way up to H$_{p}$(0) where $\chi_{s}$=$\frac{N_{1}(0)}{N_{1}(0)+N_{2}(0)}\cdot \chi_{n}$. For instance, H$_{p}$(0) would be enhanced by a factor of about 1.4 for N$_{1}$ = N$_{2}$. However, an accurate value of H$_{p}$(0) from Eq. \ref{eq8} cannot be derived at present because the Fermi-surface of KFe$_{2}$As$_{2}$ consists of 4 sheets and all the gaps and densities of states have not been determined yet. Since the large gap ultimately determined H$_{c2}$(0), an approximate value can nevertheless be obtained using the following expression
\begin{equation}
	\frac{H_{c2}(\alpha_{M})}{H_{orb}(0)} = \frac{1}{\sqrt{1+0.6\cdot\alpha_{M}^{2}}},
\label{eq9}
\end{equation}
derived by Machida and Ichioka~\cite{Ichioka07,Machida08} for a single band superconductor in the clean limit. For H $||$ $a$, we find H$_{c2}$(0)/{H$_{orb}$(0) $\approx$ 0.6 which gives $\alpha_{M}$ $\approx$ 1.8, in good agreement with our simple estimate (see Section \ref{sec:magnet} and inset of Fig.\ref{Fig2}(b)).  We note that this corresponds roughly to the minimum value that allows the formation of the FFLO phase in the presence of orbital effects,~\cite{Gruenberg66} and warrants more low temperature measurements on this system. Our estimated values for all these fields are summarized in Table \ref{table:resume}. 

\subsection{Uniaxial pressure effects}\label{sec:discussion2}    
A summary of the uniaxial pressure derivatives is given in Table \ref{table1}; here we have also included the relative pressure derivatives, which allows us to directly compare the magnitude of the various derivatives. The largest relative pressure derivatives are for $T_{c}$ and $H_{c}$, which are roughly equal both in magnitude and sign. Thus, the critical temperature and the condensation energy are strongly linked, which is not surprising. On the other hand, the relative pressure derivative of $\gamma_{n}$ and $\chi$ are much smaller than those of $T_{c}$ and $H_{c}$. Also, there is no direct correlation between the signs of $dT_{c}/dp_{i}$ and $d\gamma_{n}/dp_{i}$, which implies that these quantities are not directly related. Here it is worth pointing out that pressure and doping are strongly correlated in Co- and P-doped systems,~\cite{Meingast12,Boehmer12} which manifests itself in a similar dependence of $T_{c}$ and $\gamma_{n}$ versus either doping or pressure. Such a correlation between doping and pressure does not appear to work for the K-doped systems. Indeed, $\gamma_{n}$ is largest and $T_{c}$ is lowest for KFe$_{2}$As$_{2}$. Thus, if a similar equivalence would hold, one would expect that under hydrostatic pressure $\gamma_{n}$ would increase since $T_{c}$ decreases, which is the opposite of the observed behavior (see Table \ref{table1}). Using a bulk elastic modulus from the DFT calculation of $B$ = 45 GPa,~\cite{Heid13} we can also calculate volume Gr\"uneisen parameters of the various physical quantities (see Table \ref{table1}). Interestingly, these volume Gr\"uneisen parameters of $T_{c}$ and $\gamma_{n}$ are of similar magnitude and sign as found in various U-based heavy fermion materials,~\cite{Flouquet91} which was interpreted in terms of a negative pressure dependence of the pairing interaction. Similar physics may be operating in KFe$_{2}$As$_{2}$, which may be considered 3$d$ heavy fermion metal.~\cite{Hardy13}  

\begingroup		
\squeezetable	
\begin{table*}
\caption{Table of uniaxial pressure dependencies along the $a$- and the $c$-axis, the corresponding normalized pressure dependencies in units of GPa$^{-1}$ and the Gr\"uneisen parameters. Values of $\chi^{a}$ and $\chi^{c}$ are taken from Ref.~\cite{Hardy13}.}
\begin{ruledtabular}
\begin{tabular}{c || c | c | c | c | c }
&	$\gamma_{n}$ = 102 mJ mol$^{-1}$K$^{-2}$ & $T_{c}$ = 3.45 K& $\chi^{a}=4.1\times10^{-4}$& $\chi^{c}=3.2\times10^{-4}$& $H^{a}_{c}$ = 0.72 T \\ \hline
\hline
\multicolumn{6}{c}{}\\
\hline
\hline
& $d\gamma_{n}/dp_{i}$ [mJ mol$^{-1}$K$^{-2}$GPa$^{-1}$]& $dT_{c}/dp_{i}$ [K GPa$^{-1}$]& $d\chi^{a}/dp_{i}$ [GPa$^{-1}$]& $d\chi^{c}/dp_{i}$ [GPa$^{-1}$]& $dH^{a}_{c}/dp_{i}$ [T GPa$^{-1}$]\\ \hline
$a$-axis & $-7.74$ & $-1.92$ & $-1.43\times 10^{-5}$ &  $4.13\times 10^{-6}$ & $-0.049$ \\ 
$c$-axis & $-4.81$ &  $2.10$ & $-6.74\times 10^{-6}$ & $-3.08\times 10^{-5}$ & $0.046$ \\ \hline
\hline
\multicolumn{6}{c}{}\\
\hline
\hline			 
& $d \ln \gamma_{n}/dp_{i}$& $d\ln T_{c}/dp_{i}$& $d\ln \chi^{a}/dp_{i}$& $d\ln \chi^{c}/dp_{i}$& $d\ln H^{a}_{c}/dp_{i}$\\ \hline 
$a$-axis & $-0.076$ & $-0.56$ & $-0.035$ &  $0.013$ & $-0.709$ \\ 
$c$-axis & $-0.047$ &  $0.61$ & $-0.016$ & $-0.096$ &  $0.664$ \\ 
Volume   & $-0.199$ & $-0.51$ & $-0.086$ & $-0.070$ & $-0.754$ \\ \hline \hline
\multicolumn{6}{c}{}\\
\hline
\hline
Gr\"uneisen parameter & $d \ln \gamma_{n}/d \ln V$& $d \ln T_{c}/d \ln V$& $d \ln \chi^{a}/d \ln V$& $d \ln \chi^{c}/d \ln V$& $d \ln H^{a}_{c}/d \ln V$ \\ \hline 

Volume   & $-8.9$ &  $-22.9$ & $-3.9$ &  $-3.1$ &  $-33.9$ \\
\end{tabular}
\end{ruledtabular}
\label{table1}
\end{table*}
\endgroup		

\section{Conclusions}\label{sec:conclusions}
In conclusion, the present thermodynamic investigation of KFe$_{2}$As$_{2}$ has revealed several interesting results. Clear evidence for strong Pauli-limiting behavior is observed in the low-temperature magnetization measurements. Further detailed studies are needed to determine if the low-temperature transition is effectively first-order and for searching for other ordering phenomena, such as the elusive FFLO state. Interestingly, the derived uniaxial pressure derivatives show that superconductivity in KFe$_{2}$As$_{2}$ responds strongly to the $c/a$ ratio of the lattice constants, but with the opposite sign as in Co- and P-doped Ba122 iron pnictides, showing that this is not a universal characteristic of these materials. It is hoped that the various pressure derivatives derived here will add a stringent constraint on superconducting theories of this interesting low-$T_{c}$ compound.

\begin{acknowledgements}
The authors thank J. Schmalian, I. Vekhter, K. Machida and M. Ichioka for useful discussions. This work has been supported by the DFG through SPP1458. The work performed in Grenoble was supported by the French ANR Projects SINUS and CHIRnMAG and the ERC starting grant NewHeavyFermion.
\end{acknowledgements}

\bibliography{biblio}

\begin{thebibliography}{58}
\expandafter\ifx\csname natexlab\endcsname\relax\def\natexlab#1{#1}\fi
\expandafter\ifx\csname bibnamefont\endcsname\relax
  \def\bibnamefont#1{#1}\fi
\expandafter\ifx\csname bibfnamefont\endcsname\relax
  \def\bibfnamefont#1{#1}\fi
\expandafter\ifx\csname citenamefont\endcsname\relax
  \def\citenamefont#1{#1}\fi
\expandafter\ifx\csname url\endcsname\relax
  \def\url#1{\texttt{#1}}\fi
\expandafter\ifx\csname urlprefix\endcsname\relax\def\urlprefix{URL }\fi
\providecommand{\bibinfo}[2]{#2}
\providecommand{\eprint}[2][]{\url{#2}}

\bibitem[{\citenamefont{Johnston}(2010)}]{Johnston10}
\bibinfo{author}{\bibfnamefont{D.~C.} \bibnamefont{Johnston}},
  \bibinfo{journal}{Adv. Phys.} \textbf{\bibinfo{volume}{59}},
  \bibinfo{pages}{803} (\bibinfo{year}{2010}).

\bibitem[{\citenamefont{Paglione and Greene}(2010)}]{Paglione10}
\bibinfo{author}{\bibfnamefont{J.}~\bibnamefont{Paglione}} \bibnamefont{and}
  \bibinfo{author}{\bibfnamefont{R.~L.} \bibnamefont{Greene}},
  \bibinfo{journal}{Nat. Phys.} \textbf{\bibinfo{volume}{6}},
  \bibinfo{pages}{645} (\bibinfo{year}{2010}).

\bibitem[{\citenamefont{Stewart}(2011)}]{Stewart11}
\bibinfo{author}{\bibfnamefont{G.~R.} \bibnamefont{Stewart}},
  \bibinfo{journal}{Rev. Mod. Phys.} \textbf{\bibinfo{volume}{83}},
  \bibinfo{pages}{1589} (\bibinfo{year}{2011}).

\bibitem[{\citenamefont{Mazin and Schmalian}(2009)}]{Mazin09}
\bibinfo{author}{\bibfnamefont{I.~I.} \bibnamefont{Mazin}} \bibnamefont{and}
  \bibinfo{author}{\bibfnamefont{J.}~\bibnamefont{Schmalian}},
  \bibinfo{journal}{Physica C} \textbf{\bibinfo{volume}{469}},
  \bibinfo{pages}{614} (\bibinfo{year}{2009}).

\bibitem[{\citenamefont{Fukazawa et~al.}(2009)\citenamefont{Fukazawa, Yamada,
  Kondo, Saito, Kohori, Kuga, Matsumoto, Nakatsuji, Kito, Shirage
  et~al.}}]{Fukazawa09}
\bibinfo{author}{\bibfnamefont{H.}~\bibnamefont{Fukazawa}},
  \bibinfo{author}{\bibfnamefont{Y.}~\bibnamefont{Yamada}},
  \bibinfo{author}{\bibfnamefont{K.}~\bibnamefont{Kondo}},
  \bibinfo{author}{\bibfnamefont{T.}~\bibnamefont{Saito}},
  \bibinfo{author}{\bibfnamefont{Y.}~\bibnamefont{Kohori}},
  \bibinfo{author}{\bibfnamefont{K.}~\bibnamefont{Kuga}},
  \bibinfo{author}{\bibfnamefont{Y.}~\bibnamefont{Matsumoto}},
  \bibinfo{author}{\bibfnamefont{S.}~\bibnamefont{Nakatsuji}},
  \bibinfo{author}{\bibfnamefont{H.}~\bibnamefont{Kito}},
  \bibinfo{author}{\bibfnamefont{P.~M.} \bibnamefont{Shirage}},
  \bibnamefont{et~al.}, \bibinfo{journal}{J. Phys. Soc. Jpn}
  \textbf{\bibinfo{volume}{78}}, \bibinfo{pages}{083712}
  (\bibinfo{year}{2009}).

\bibitem[{\citenamefont{Terashima et~al.}(2009)\citenamefont{Terashima, Kimata,
  Satsukawa, Harada, Hazama, Uji, Harima, Chen, Luo, and Wang}}]{Terashima09}
\bibinfo{author}{\bibfnamefont{T.}~\bibnamefont{Terashima}},
  \bibinfo{author}{\bibfnamefont{M.}~\bibnamefont{Kimata}},
  \bibinfo{author}{\bibfnamefont{H.}~\bibnamefont{Satsukawa}},
  \bibinfo{author}{\bibfnamefont{A.}~\bibnamefont{Harada}},
  \bibinfo{author}{\bibfnamefont{K.}~\bibnamefont{Hazama}},
  \bibinfo{author}{\bibfnamefont{S.}~\bibnamefont{Uji}},
  \bibinfo{author}{\bibfnamefont{H.}~\bibnamefont{Harima}},
  \bibinfo{author}{\bibfnamefont{G.-F.} \bibnamefont{Chen}},
  \bibinfo{author}{\bibfnamefont{J.-L.} \bibnamefont{Luo}}, \bibnamefont{and}
  \bibinfo{author}{\bibfnamefont{N.-L.} \bibnamefont{Wang}},
  \bibinfo{journal}{J. Phys. Soc. Jpn} \textbf{\bibinfo{volume}{78}},
  \bibinfo{pages}{063702} (\bibinfo{year}{2009}).

\bibitem[{\citenamefont{Rotter et~al.}(2008)\citenamefont{Rotter, Tegel, and
  Johrendt}}]{Rotter08}
\bibinfo{author}{\bibfnamefont{M.}~\bibnamefont{Rotter}},
  \bibinfo{author}{\bibfnamefont{M.}~\bibnamefont{Tegel}}, \bibnamefont{and}
  \bibinfo{author}{\bibfnamefont{D.}~\bibnamefont{Johrendt}},
  \bibinfo{journal}{Phys. Rev. Lett.} \textbf{\bibinfo{volume}{101}},
  \bibinfo{pages}{107006} (\bibinfo{year}{2008}).

\bibitem[{\citenamefont{Popovich et~al.}(2010)\citenamefont{Popovich, Boris,
  Dolgov, Golubov, Sun, Lin, Kremer, and Keimer}}]{Popovich10}
\bibinfo{author}{\bibfnamefont{P.}~\bibnamefont{Popovich}},
  \bibinfo{author}{\bibfnamefont{A.~V.} \bibnamefont{Boris}},
  \bibinfo{author}{\bibfnamefont{O.~V.} \bibnamefont{Dolgov}},
  \bibinfo{author}{\bibfnamefont{A.~A.} \bibnamefont{Golubov}},
  \bibinfo{author}{\bibfnamefont{D.~L.} \bibnamefont{Sun}},
  \bibinfo{author}{\bibfnamefont{C.~T.} \bibnamefont{Lin}},
  \bibinfo{author}{\bibfnamefont{R.~K.} \bibnamefont{Kremer}},
  \bibnamefont{and} \bibinfo{author}{\bibfnamefont{B.}~\bibnamefont{Keimer}},
  \bibinfo{journal}{Phys. Rev. Lett.} \textbf{\bibinfo{volume}{105}},
  \bibinfo{pages}{027003} (\bibinfo{year}{2010}).

\bibitem[{\citenamefont{Li et~al.}(2011)\citenamefont{Li, Sun, Lin, Su, Hu, and
  Zheng}}]{Li11}
\bibinfo{author}{\bibfnamefont{Z.}~\bibnamefont{Li}},
  \bibinfo{author}{\bibfnamefont{D.~L.} \bibnamefont{Sun}},
  \bibinfo{author}{\bibfnamefont{C.~T.} \bibnamefont{Lin}},
  \bibinfo{author}{\bibfnamefont{Y.~H.} \bibnamefont{Su}},
  \bibinfo{author}{\bibfnamefont{J.~P.} \bibnamefont{Hu}}, \bibnamefont{and}
  \bibinfo{author}{\bibfnamefont{G.~Q.} \bibnamefont{Zheng}},
  \bibinfo{journal}{Phys. Rev. B} \textbf{\bibinfo{volume}{83}},
  \bibinfo{pages}{140506} (\bibinfo{year}{2011}).

\bibitem[{\citenamefont{Thomale et~al.}(2011)\citenamefont{Thomale, Platt,
  Hanke, Hu, and Bernevig}}]{Thomale11}
\bibinfo{author}{\bibfnamefont{R.}~\bibnamefont{Thomale}},
  \bibinfo{author}{\bibfnamefont{C.}~\bibnamefont{Platt}},
  \bibinfo{author}{\bibfnamefont{W.}~\bibnamefont{Hanke}},
  \bibinfo{author}{\bibfnamefont{J.}~\bibnamefont{Hu}}, \bibnamefont{and}
  \bibinfo{author}{\bibfnamefont{B.~A.} \bibnamefont{Bernevig}},
  \bibinfo{journal}{Phys. Rev. Lett.} \textbf{\bibinfo{volume}{107}},
  \bibinfo{pages}{117001} (\bibinfo{year}{2011}).

\bibitem[{\citenamefont{Hashimoto et~al.}(2010)\citenamefont{Hashimoto,
  Serafin, Tonegawa, Katsumata, Okazaki, Saito, Fukazawa, Kohori, Kihou, Lee
  et~al.}}]{Hashimoto10}
\bibinfo{author}{\bibfnamefont{K.}~\bibnamefont{Hashimoto}},
  \bibinfo{author}{\bibfnamefont{A.}~\bibnamefont{Serafin}},
  \bibinfo{author}{\bibfnamefont{S.}~\bibnamefont{Tonegawa}},
  \bibinfo{author}{\bibfnamefont{R.}~\bibnamefont{Katsumata}},
  \bibinfo{author}{\bibfnamefont{R.}~\bibnamefont{Okazaki}},
  \bibinfo{author}{\bibfnamefont{T.}~\bibnamefont{Saito}},
  \bibinfo{author}{\bibfnamefont{H.}~\bibnamefont{Fukazawa}},
  \bibinfo{author}{\bibfnamefont{Y.}~\bibnamefont{Kohori}},
  \bibinfo{author}{\bibfnamefont{K.}~\bibnamefont{Kihou}},
  \bibinfo{author}{\bibfnamefont{C.~H.} \bibnamefont{Lee}},
  \bibnamefont{et~al.}, \bibinfo{journal}{Phys. Rev. B}
  \textbf{\bibinfo{volume}{82}}, \bibinfo{pages}{014526}
  (\bibinfo{year}{2010}).

\bibitem[{\citenamefont{Reid et~al.}(2012)\citenamefont{Reid, Tanatar,
  Juneau-Fecteau, Gordon, de~Cotret, Doiron-Leyraud, Saito, Fukazawa, Kohori,
  Kihou et~al.}}]{Reid12}
\bibinfo{author}{\bibfnamefont{J.-P.} \bibnamefont{Reid}},
  \bibinfo{author}{\bibfnamefont{M.~A.} \bibnamefont{Tanatar}},
  \bibinfo{author}{\bibfnamefont{A.}~\bibnamefont{Juneau-Fecteau}},
  \bibinfo{author}{\bibfnamefont{R.~T.} \bibnamefont{Gordon}},
  \bibinfo{author}{\bibfnamefont{S.~R.} \bibnamefont{de~Cotret}},
  \bibinfo{author}{\bibfnamefont{N.}~\bibnamefont{Doiron-Leyraud}},
  \bibinfo{author}{\bibfnamefont{T.}~\bibnamefont{Saito}},
  \bibinfo{author}{\bibfnamefont{H.}~\bibnamefont{Fukazawa}},
  \bibinfo{author}{\bibfnamefont{Y.}~\bibnamefont{Kohori}},
  \bibinfo{author}{\bibfnamefont{K.}~\bibnamefont{Kihou}},
  \bibnamefont{et~al.}, \bibinfo{journal}{Phys. Rev. Lett.}
  \textbf{\bibinfo{volume}{109}}, \bibinfo{pages}{087001}
  (\bibinfo{year}{2012}).

\bibitem[{\citenamefont{Okazaki et~al.}(2012)\citenamefont{Okazaki, Ota,
  Kotani, Malaeb, Ishida, Shimojima, Kiss, Watanabe, Chen, Kihou
  et~al.}}]{Okazaki12}
\bibinfo{author}{\bibfnamefont{K.}~\bibnamefont{Okazaki}},
  \bibinfo{author}{\bibfnamefont{Y.}~\bibnamefont{Ota}},
  \bibinfo{author}{\bibfnamefont{Y.}~\bibnamefont{Kotani}},
  \bibinfo{author}{\bibfnamefont{W.}~\bibnamefont{Malaeb}},
  \bibinfo{author}{\bibfnamefont{Y.}~\bibnamefont{Ishida}},
  \bibinfo{author}{\bibfnamefont{T.}~\bibnamefont{Shimojima}},
  \bibinfo{author}{\bibfnamefont{T.}~\bibnamefont{Kiss}},
  \bibinfo{author}{\bibfnamefont{S.}~\bibnamefont{Watanabe}},
  \bibinfo{author}{\bibfnamefont{C.-T.} \bibnamefont{Chen}},
  \bibinfo{author}{\bibfnamefont{K.}~\bibnamefont{Kihou}},
  \bibnamefont{et~al.}, \bibinfo{journal}{Science}
  \textbf{\bibinfo{volume}{337}}, \bibinfo{pages}{1314} (\bibinfo{year}{2012}).

\bibitem[{\citenamefont{Sato et~al.}(2009)\citenamefont{Sato, Nakayama, Sekiba,
  Richard, Y.-M.Xu, Souma, Takahashi, Chen, Luo, Wang et~al.}}]{Sato09}
\bibinfo{author}{\bibfnamefont{T.}~\bibnamefont{Sato}},
  \bibinfo{author}{\bibfnamefont{K.}~\bibnamefont{Nakayama}},
  \bibinfo{author}{\bibfnamefont{Y.}~\bibnamefont{Sekiba}},
  \bibinfo{author}{\bibfnamefont{P.}~\bibnamefont{Richard}},
  \bibinfo{author}{\bibnamefont{Y.-M.Xu}},
  \bibinfo{author}{\bibfnamefont{S.}~\bibnamefont{Souma}},
  \bibinfo{author}{\bibfnamefont{T.}~\bibnamefont{Takahashi}},
  \bibinfo{author}{\bibfnamefont{G.~F.} \bibnamefont{Chen}},
  \bibinfo{author}{\bibfnamefont{J.~L.} \bibnamefont{Luo}},
  \bibinfo{author}{\bibfnamefont{N.~L.} \bibnamefont{Wang}},
  \bibnamefont{et~al.}, \bibinfo{journal}{Phys. Rev. Lett.}
  \textbf{\bibinfo{volume}{103}}, \bibinfo{pages}{047002}
  (\bibinfo{year}{2009}).

\bibitem[{\citenamefont{Lee et~al.}(2011)\citenamefont{Lee, Kihou,
  Kawano-Furukawa, Saito, Iyo, Eisaki, Fukazawa, Kohori, Suzuki, Usui
  et~al.}}]{Lee11}
\bibinfo{author}{\bibfnamefont{C.~H.} \bibnamefont{Lee}},
  \bibinfo{author}{\bibfnamefont{K.}~\bibnamefont{Kihou}},
  \bibinfo{author}{\bibfnamefont{H.}~\bibnamefont{Kawano-Furukawa}},
  \bibinfo{author}{\bibfnamefont{T.}~\bibnamefont{Saito}},
  \bibinfo{author}{\bibfnamefont{A.}~\bibnamefont{Iyo}},
  \bibinfo{author}{\bibfnamefont{H.}~\bibnamefont{Eisaki}},
  \bibinfo{author}{\bibfnamefont{H.}~\bibnamefont{Fukazawa}},
  \bibinfo{author}{\bibfnamefont{Y.}~\bibnamefont{Kohori}},
  \bibinfo{author}{\bibfnamefont{K.}~\bibnamefont{Suzuki}},
  \bibinfo{author}{\bibfnamefont{H.}~\bibnamefont{Usui}}, \bibnamefont{et~al.},
  \bibinfo{journal}{Phys. Rev. Lett.} \textbf{\bibinfo{volume}{106}},
  \bibinfo{pages}{067003} (\bibinfo{year}{2011}).

\bibitem[{\citenamefont{Castellan et~al.}(2011)\citenamefont{Castellan,
  Rosenkranz, Goremychkin, Chung, Todorov, Kanatzidis, Eremin, Knolle,
  Chubukov, Maiti et~al.}}]{Castellan11}
\bibinfo{author}{\bibfnamefont{J.-P.} \bibnamefont{Castellan}},
  \bibinfo{author}{\bibfnamefont{S.}~\bibnamefont{Rosenkranz}},
  \bibinfo{author}{\bibfnamefont{E.~A.} \bibnamefont{Goremychkin}},
  \bibinfo{author}{\bibfnamefont{D.~Y.} \bibnamefont{Chung}},
  \bibinfo{author}{\bibfnamefont{I.~S.} \bibnamefont{Todorov}},
  \bibinfo{author}{\bibfnamefont{M.~G.} \bibnamefont{Kanatzidis}},
  \bibinfo{author}{\bibfnamefont{I.}~\bibnamefont{Eremin}},
  \bibinfo{author}{\bibfnamefont{J.}~\bibnamefont{Knolle}},
  \bibinfo{author}{\bibfnamefont{A.~V.} \bibnamefont{Chubukov}},
  \bibinfo{author}{\bibfnamefont{S.}~\bibnamefont{Maiti}},
  \bibnamefont{et~al.}, \bibinfo{journal}{Phys. Rev. Lett.}
  \textbf{\bibinfo{volume}{107}}, \bibinfo{pages}{177003}
  (\bibinfo{year}{2011}).

\bibitem[{\citenamefont{Hardy et~al.}(2013)\citenamefont{Hardy, B\"ohmer, Aoki,
  Burger, Wolf, Schweiss, Heid, Adelmann, Yao, Kotliar et~al.}}]{Hardy13}
\bibinfo{author}{\bibfnamefont{F.}~\bibnamefont{Hardy}},
  \bibinfo{author}{\bibfnamefont{A.}~\bibnamefont{B\"ohmer}},
  \bibinfo{author}{\bibfnamefont{D.}~\bibnamefont{Aoki}},
  \bibinfo{author}{\bibfnamefont{P.}~\bibnamefont{Burger}},
  \bibinfo{author}{\bibfnamefont{T.}~\bibnamefont{Wolf}},
  \bibinfo{author}{\bibfnamefont{P.}~\bibnamefont{Schweiss}},
  \bibinfo{author}{\bibfnamefont{R.}~\bibnamefont{Heid}},
  \bibinfo{author}{\bibfnamefont{P.}~\bibnamefont{Adelmann}},
  \bibinfo{author}{\bibfnamefont{Y.}~\bibnamefont{Yao}},
  \bibinfo{author}{\bibfnamefont{G.}~\bibnamefont{Kotliar}},
  \bibnamefont{et~al.}, \bibinfo{journal}{ArXiv e-prints}
  \textbf{\bibinfo{volume}{1302.1696}} (\bibinfo{year}{2013}).

\bibitem[{\citenamefont{de' Medici et~al.}(2012)\citenamefont{de' Medici,
  Giovannetti, and Capone}}]{deMedici12}
\bibinfo{author}{\bibfnamefont{L.}~\bibnamefont{de' Medici}},
  \bibinfo{author}{\bibfnamefont{G.}~\bibnamefont{Giovannetti}},
  \bibnamefont{and} \bibinfo{author}{\bibfnamefont{M.}~\bibnamefont{Capone}},
  \bibinfo{journal}{ArXiv e-prints} \textbf{\bibinfo{volume}{1212.3966}}
  (\bibinfo{year}{2012}).

\bibitem[{\citenamefont{P.~Fulde and Ferrell}(1964)}]{Fulde64}
\bibinfo{author}{\bibfnamefont{P.}~\bibnamefont{P.~Fulde}} \bibnamefont{and}
  \bibinfo{author}{\bibfnamefont{R.~A.} \bibnamefont{Ferrell}},
  \bibinfo{journal}{Phys. Rev.} \textbf{\bibinfo{volume}{135}},
  \bibinfo{pages}{A550} (\bibinfo{year}{1964}).

\bibitem[{\citenamefont{Larkin and Ovchinnikov}(1965)}]{Larkin65}
\bibinfo{author}{\bibfnamefont{A.~I.} \bibnamefont{Larkin}} \bibnamefont{and}
  \bibinfo{author}{\bibfnamefont{Y.~N.} \bibnamefont{Ovchinnikov}},
  \bibinfo{journal}{Sov. Phys. JETP} \textbf{\bibinfo{volume}{20}},
  \bibinfo{pages}{762} (\bibinfo{year}{1965}).

\bibitem[{\citenamefont{Hardy et~al.}(2009)\citenamefont{Hardy, Adelmann, Wolf,
  v.~L\"ohneysen, and Meingast}}]{Hardy09}
\bibinfo{author}{\bibfnamefont{F.}~\bibnamefont{Hardy}},
  \bibinfo{author}{\bibfnamefont{P.}~\bibnamefont{Adelmann}},
  \bibinfo{author}{\bibfnamefont{T.}~\bibnamefont{Wolf}},
  \bibinfo{author}{\bibfnamefont{H.}~\bibnamefont{v.~L\"ohneysen}},
  \bibnamefont{and} \bibinfo{author}{\bibfnamefont{C.}~\bibnamefont{Meingast}},
  \bibinfo{journal}{Phys. Rev. Lett.} \textbf{\bibinfo{volume}{102}},
  \bibinfo{pages}{187004} (\bibinfo{year}{2009}).

\bibitem[{\citenamefont{Meingast et~al.}(2012)\citenamefont{Meingast, Hardy,
  Heid, Adelmann, B\"ohmer, Burger, Ernst, Fromknecht, Schweiss, and
  Wolf}}]{Meingast12}
\bibinfo{author}{\bibfnamefont{C.}~\bibnamefont{Meingast}},
  \bibinfo{author}{\bibfnamefont{F.}~\bibnamefont{Hardy}},
  \bibinfo{author}{\bibfnamefont{R.}~\bibnamefont{Heid}},
  \bibinfo{author}{\bibfnamefont{P.}~\bibnamefont{Adelmann}},
  \bibinfo{author}{\bibfnamefont{A.}~\bibnamefont{B\"ohmer}},
  \bibinfo{author}{\bibfnamefont{P.}~\bibnamefont{Burger}},
  \bibinfo{author}{\bibfnamefont{D.}~\bibnamefont{Ernst}},
  \bibinfo{author}{\bibfnamefont{R.}~\bibnamefont{Fromknecht}},
  \bibinfo{author}{\bibfnamefont{P.}~\bibnamefont{Schweiss}}, \bibnamefont{and}
  \bibinfo{author}{\bibfnamefont{T.}~\bibnamefont{Wolf}},
  \bibinfo{journal}{Phys. Rev. Lett.} \textbf{\bibinfo{volume}{108}},
  \bibinfo{pages}{177004} (\bibinfo{year}{2012}).

\bibitem[{\citenamefont{B\"ohmer et~al.}(2012)\citenamefont{B\"ohmer, Burger,
  Hardy, Wolf, Schweiss, Fromknecht, v.~L\"ohneysen, Meingast, Mak, Lortz
  et~al.}}]{Boehmer12}
\bibinfo{author}{\bibfnamefont{A.~E.} \bibnamefont{B\"ohmer}},
  \bibinfo{author}{\bibfnamefont{P.}~\bibnamefont{Burger}},
  \bibinfo{author}{\bibfnamefont{F.}~\bibnamefont{Hardy}},
  \bibinfo{author}{\bibfnamefont{T.}~\bibnamefont{Wolf}},
  \bibinfo{author}{\bibfnamefont{P.}~\bibnamefont{Schweiss}},
  \bibinfo{author}{\bibfnamefont{R.}~\bibnamefont{Fromknecht}},
  \bibinfo{author}{\bibfnamefont{H.}~\bibnamefont{v.~L\"ohneysen}},
  \bibinfo{author}{\bibfnamefont{C.}~\bibnamefont{Meingast}},
  \bibinfo{author}{\bibfnamefont{H.~K.} \bibnamefont{Mak}},
  \bibinfo{author}{\bibfnamefont{R.}~\bibnamefont{Lortz}},
  \bibnamefont{et~al.}, \bibinfo{journal}{Phys. Rev. B}
  \textbf{\bibinfo{volume}{86}}, \bibinfo{pages}{094521}
  (\bibinfo{year}{2012}).

\bibitem[{\citenamefont{Meingast et~al.}(1990)\citenamefont{Meingast, Blank,
  B\"urkle, Obst, Wolf, W\"uhl, Selvamanickam, and Salama}}]{Meingast90}
\bibinfo{author}{\bibfnamefont{C.}~\bibnamefont{Meingast}},
  \bibinfo{author}{\bibfnamefont{B.}~\bibnamefont{Blank}},
  \bibinfo{author}{\bibfnamefont{H.}~\bibnamefont{B\"urkle}},
  \bibinfo{author}{\bibfnamefont{B.}~\bibnamefont{Obst}},
  \bibinfo{author}{\bibfnamefont{T.}~\bibnamefont{Wolf}},
  \bibinfo{author}{\bibfnamefont{H.}~\bibnamefont{W\"uhl}},
  \bibinfo{author}{\bibfnamefont{V.}~\bibnamefont{Selvamanickam}},
  \bibnamefont{and} \bibinfo{author}{\bibfnamefont{K.}~\bibnamefont{Salama}},
  \bibinfo{journal}{Phys. Rev. B} \textbf{\bibinfo{volume}{41}},
  \bibinfo{pages}{11299} (\bibinfo{year}{1990}).

\bibitem[{\citenamefont{Grinenko et~al.}(2013)\citenamefont{Grinenko,
  Drechsler, Abdel-Hafiez, Aswartham, Wolter, Wurmehl, Hess, Nenkov, Fuchs,
  Efremov et~al.}}]{Grinenko13}
\bibinfo{author}{\bibfnamefont{V.}~\bibnamefont{Grinenko}},
  \bibinfo{author}{\bibfnamefont{S.-L.} \bibnamefont{Drechsler}},
  \bibinfo{author}{\bibfnamefont{M.}~\bibnamefont{Abdel-Hafiez}},
  \bibinfo{author}{\bibfnamefont{S.}~\bibnamefont{Aswartham}},
  \bibinfo{author}{\bibfnamefont{A.~U.~B.} \bibnamefont{Wolter}},
  \bibinfo{author}{\bibfnamefont{S.}~\bibnamefont{Wurmehl}},
  \bibinfo{author}{\bibfnamefont{C.}~\bibnamefont{Hess}},
  \bibinfo{author}{\bibfnamefont{K.}~\bibnamefont{Nenkov}},
  \bibinfo{author}{\bibfnamefont{G.}~\bibnamefont{Fuchs}},
  \bibinfo{author}{\bibfnamefont{D.~V.} \bibnamefont{Efremov}},
  \bibnamefont{et~al.}, \bibinfo{journal}{Phys. Status Solidi B}
  \textbf{\bibinfo{volume}{250}}, \bibinfo{pages}{593} (\bibinfo{year}{2013}).

\bibitem[{\citenamefont{Kawano-Furukawa
  et~al.}(2011)\citenamefont{Kawano-Furukawa, Bowell, White, Heslop, Cameron,
  Forgan, Kihou, Lee, Iyo, Eisaki et~al.}}]{Kawano11}
\bibinfo{author}{\bibfnamefont{H.}~\bibnamefont{Kawano-Furukawa}},
  \bibinfo{author}{\bibfnamefont{C.~J.} \bibnamefont{Bowell}},
  \bibinfo{author}{\bibfnamefont{J.~S.} \bibnamefont{White}},
  \bibinfo{author}{\bibfnamefont{R.~W.} \bibnamefont{Heslop}},
  \bibinfo{author}{\bibfnamefont{A.~S.} \bibnamefont{Cameron}},
  \bibinfo{author}{\bibfnamefont{E.~M.} \bibnamefont{Forgan}},
  \bibinfo{author}{\bibfnamefont{K.}~\bibnamefont{Kihou}},
  \bibinfo{author}{\bibfnamefont{C.~H.} \bibnamefont{Lee}},
  \bibinfo{author}{\bibfnamefont{A.}~\bibnamefont{Iyo}},
  \bibinfo{author}{\bibfnamefont{H.}~\bibnamefont{Eisaki}},
  \bibnamefont{et~al.}, \bibinfo{journal}{Phys. Rev. B}
  \textbf{\bibinfo{volume}{84}}, \bibinfo{pages}{024507}
  (\bibinfo{year}{2011}).

\bibitem[{\citenamefont{Eskildsen et~al.}(2009)\citenamefont{Eskildsen,
  Vinnikov, Blasius, Veshchunov, Artemova, Densmore, Dewhurst, Ni, Kreyssig,
  Bud'ko et~al.}}]{Eskildsen09}
\bibinfo{author}{\bibfnamefont{M.~R.} \bibnamefont{Eskildsen}},
  \bibinfo{author}{\bibfnamefont{L.~Y.} \bibnamefont{Vinnikov}},
  \bibinfo{author}{\bibfnamefont{T.~D.} \bibnamefont{Blasius}},
  \bibinfo{author}{\bibfnamefont{I.~S.} \bibnamefont{Veshchunov}},
  \bibinfo{author}{\bibfnamefont{T.~M.} \bibnamefont{Artemova}},
  \bibinfo{author}{\bibfnamefont{J.~M.} \bibnamefont{Densmore}},
  \bibinfo{author}{\bibfnamefont{C.~D.} \bibnamefont{Dewhurst}},
  \bibinfo{author}{\bibfnamefont{N.}~\bibnamefont{Ni}},
  \bibinfo{author}{\bibfnamefont{A.}~\bibnamefont{Kreyssig}},
  \bibinfo{author}{\bibfnamefont{S.~L.} \bibnamefont{Bud'ko}},
  \bibnamefont{et~al.}, \bibinfo{journal}{Phys. Rev. B}
  \textbf{\bibinfo{volume}{79}}, \bibinfo{pages}{100501}
  (\bibinfo{year}{2009}).

\bibitem[{\citenamefont{Yamamoto et~al.}(2009)\citenamefont{Yamamoto,
  Jaroszynski, Tarantini, Balicas, Jiang, Gurevich, Larbalestier, Jin, Sefat,
  McGuire et~al.}}]{Yamamoto09}
\bibinfo{author}{\bibfnamefont{A.}~\bibnamefont{Yamamoto}},
  \bibinfo{author}{\bibfnamefont{J.}~\bibnamefont{Jaroszynski}},
  \bibinfo{author}{\bibfnamefont{C.}~\bibnamefont{Tarantini}},
  \bibinfo{author}{\bibfnamefont{L.}~\bibnamefont{Balicas}},
  \bibinfo{author}{\bibfnamefont{J.}~\bibnamefont{Jiang}},
  \bibinfo{author}{\bibfnamefont{A.}~\bibnamefont{Gurevich}},
  \bibinfo{author}{\bibfnamefont{D.~C.} \bibnamefont{Larbalestier}},
  \bibinfo{author}{\bibfnamefont{R.}~\bibnamefont{Jin}},
  \bibinfo{author}{\bibfnamefont{A.~S.} \bibnamefont{Sefat}},
  \bibinfo{author}{\bibfnamefont{M.~A.} \bibnamefont{McGuire}},
  \bibnamefont{et~al.}, \bibinfo{journal}{Appl. Phys. Lett.}
  \textbf{\bibinfo{volume}{94}}, \bibinfo{pages}{062511}
  (\bibinfo{year}{2009}).

\bibitem[{\citenamefont{Tayama et~al.}(2002)\citenamefont{Tayama, Harita,
  Sakakibara, Haga, Shishido, Settai, and Onuki}}]{Tayama02}
\bibinfo{author}{\bibfnamefont{T.}~\bibnamefont{Tayama}},
  \bibinfo{author}{\bibfnamefont{A.}~\bibnamefont{Harita}},
  \bibinfo{author}{\bibfnamefont{T.}~\bibnamefont{Sakakibara}},
  \bibinfo{author}{\bibfnamefont{Y.}~\bibnamefont{Haga}},
  \bibinfo{author}{\bibfnamefont{H.}~\bibnamefont{Shishido}},
  \bibinfo{author}{\bibfnamefont{R.}~\bibnamefont{Settai}}, \bibnamefont{and}
  \bibinfo{author}{\bibfnamefont{Y.}~\bibnamefont{Onuki}}, \bibinfo{journal}{J.
  Phys. Chem. Solids} \textbf{\bibinfo{volume}{63}}, \bibinfo{pages}{1155}
  (\bibinfo{year}{2002}).

\bibitem[{\citenamefont{Paulsen et~al.}(2011)\citenamefont{Paulsen, Aoki,
  Knebel, and Flouquet}}]{Paulsen11}
\bibinfo{author}{\bibfnamefont{C.}~\bibnamefont{Paulsen}},
  \bibinfo{author}{\bibfnamefont{D.}~\bibnamefont{Aoki}},
  \bibinfo{author}{\bibfnamefont{G.}~\bibnamefont{Knebel}}, \bibnamefont{and}
  \bibinfo{author}{\bibfnamefont{J.}~\bibnamefont{Flouquet}},
  \bibinfo{journal}{J. Phys. Soc. Jpn.} \textbf{\bibinfo{volume}{80}},
  \bibinfo{pages}{053701} (\bibinfo{year}{2011}).

\bibitem[{\citenamefont{Tenya et~al.}(2006)\citenamefont{Tenya, Yasuda,
  Yokoyama, Amitsuka, Deguchi, and Maeno}}]{Tenya06}
\bibinfo{author}{\bibfnamefont{K.}~\bibnamefont{Tenya}},
  \bibinfo{author}{\bibfnamefont{S.}~\bibnamefont{Yasuda}},
  \bibinfo{author}{\bibfnamefont{M.}~\bibnamefont{Yokoyama}},
  \bibinfo{author}{\bibfnamefont{H.}~\bibnamefont{Amitsuka}},
  \bibinfo{author}{\bibfnamefont{K.}~\bibnamefont{Deguchi}}, \bibnamefont{and}
  \bibinfo{author}{\bibfnamefont{Y.}~\bibnamefont{Maeno}},
  \bibinfo{journal}{Physica B} \textbf{\bibinfo{volume}{378}},
  \bibinfo{pages}{495} (\bibinfo{year}{2006}).

\bibitem[{\citenamefont{Hake}(1967)}]{Hake67}
\bibinfo{author}{\bibfnamefont{R.~R.} \bibnamefont{Hake}},
  \bibinfo{journal}{Phys. Rev.} \textbf{\bibinfo{volume}{158}},
  \bibinfo{pages}{356} (\bibinfo{year}{1967}).

\bibitem[{\citenamefont{Ichioka et~al.}(2011)\citenamefont{Ichioka, Suzuki,
  Tsutsumi, and Machida}}]{IchiokaBook}
\bibinfo{author}{\bibfnamefont{M.}~\bibnamefont{Ichioka}},
  \bibinfo{author}{\bibfnamefont{K.~M.} \bibnamefont{Suzuki}},
  \bibinfo{author}{\bibfnamefont{Y.}~\bibnamefont{Tsutsumi}}, \bibnamefont{and}
  \bibinfo{author}{\bibfnamefont{K.}~\bibnamefont{Machida}},
  \emph{\bibinfo{title}{Superconductivity - Theory and Applications}}
  (\bibinfo{publisher}{In-Tech}, \bibinfo{year}{2011}),
  chap.~\bibinfo{chapter}{10}.

\bibitem[{\citenamefont{Ichioka and Machida}(2007)}]{Ichioka07}
\bibinfo{author}{\bibfnamefont{M.}~\bibnamefont{Ichioka}} \bibnamefont{and}
  \bibinfo{author}{\bibfnamefont{K.}~\bibnamefont{Machida}},
  \bibinfo{journal}{Phys. Rev. B} \textbf{\bibinfo{volume}{76}},
  \bibinfo{pages}{064502} (\bibinfo{year}{2007}).

\bibitem[{\citenamefont{Maki}(1964)}]{Maki64}
\bibinfo{author}{\bibfnamefont{K.}~\bibnamefont{Maki}},
  \bibinfo{journal}{Physics} \textbf{\bibinfo{volume}{1}}, \bibinfo{pages}{127}
  (\bibinfo{year}{1964}).

\bibitem[{\citenamefont{Abdel-Hafiez et~al.}(2012)\citenamefont{Abdel-Hafiez,
  Aswartham, Wurmehl, Grinenko, Hess, Drechsler, Johnston, Wolter, B\"uchner,
  Rosner et~al.}}]{Hafiez12}
\bibinfo{author}{\bibfnamefont{M.}~\bibnamefont{Abdel-Hafiez}},
  \bibinfo{author}{\bibfnamefont{S.}~\bibnamefont{Aswartham}},
  \bibinfo{author}{\bibfnamefont{S.}~\bibnamefont{Wurmehl}},
  \bibinfo{author}{\bibfnamefont{V.}~\bibnamefont{Grinenko}},
  \bibinfo{author}{\bibfnamefont{C.}~\bibnamefont{Hess}},
  \bibinfo{author}{\bibfnamefont{S.~L.} \bibnamefont{Drechsler}},
  \bibinfo{author}{\bibfnamefont{S.}~\bibnamefont{Johnston}},
  \bibinfo{author}{\bibfnamefont{A.~U.~B.} \bibnamefont{Wolter}},
  \bibinfo{author}{\bibfnamefont{B.}~\bibnamefont{B\"uchner}},
  \bibinfo{author}{\bibfnamefont{H.}~\bibnamefont{Rosner}},
  \bibnamefont{et~al.}, \bibinfo{journal}{Phys. Rev. B}
  \textbf{\bibinfo{volume}{85}}, \bibinfo{pages}{134533}
  (\bibinfo{year}{2012}).

\bibitem[{\citenamefont{Ehrenfest}(1938)}]{Ehrenfest38}
\bibinfo{author}{\bibfnamefont{P.}~\bibnamefont{Ehrenfest}},
  \bibinfo{journal}{Mitteilungen aus dem Kammerlingh Onnes-Institut Leiden}
  \textbf{\bibinfo{volume}{75b}}, \bibinfo{pages}{628} (\bibinfo{year}{1938}).

\bibitem[{\citenamefont{Bud'ko et~al.}(2012)\citenamefont{Bud'ko, Liu,
  Lograsso, and Canfield}}]{Budko12}
\bibinfo{author}{\bibfnamefont{S.~L.} \bibnamefont{Bud'ko}},
  \bibinfo{author}{\bibfnamefont{Y.}~\bibnamefont{Liu}},
  \bibinfo{author}{\bibfnamefont{T.~A.} \bibnamefont{Lograsso}},
  \bibnamefont{and} \bibinfo{author}{\bibfnamefont{P.~C.}
  \bibnamefont{Canfield}}, \bibinfo{journal}{Phys. Rev. B}
  \textbf{\bibinfo{volume}{86}}, \bibinfo{pages}{224514}
  (\bibinfo{year}{2012}).

\bibitem[{\citenamefont{Shoenberg}(1962)}]{Shoenberg1962}
\bibinfo{author}{\bibfnamefont{D.}~\bibnamefont{Shoenberg}},
  \emph{\bibinfo{title}{Superconductivity}} (\bibinfo{publisher}{Cambridge
  University Press}, \bibinfo{year}{1962}).

\bibitem[{\citenamefont{Br\"andli and Enck}(1968)}]{Braendli68}
\bibinfo{author}{\bibfnamefont{G.}~\bibnamefont{Br\"andli}} \bibnamefont{and}
  \bibinfo{author}{\bibfnamefont{F.~D.} \bibnamefont{Enck}},
  \bibinfo{journal}{Phys. Lett. A} \textbf{\bibinfo{volume}{26}},
  \bibinfo{pages}{360} (\bibinfo{year}{1968}).

\bibitem[{\citenamefont{Popovych}(2006)}]{Popovych06}
\bibinfo{author}{\bibfnamefont{P.}~\bibnamefont{Popovych}}, Ph.D. thesis,
  \bibinfo{school}{Universit\"at Karlsruhe (TH)} (\bibinfo{year}{2006}).

\bibitem[{\citenamefont{Zocco et~al.}(2013)\citenamefont{Zocco, Grube, Eilers,
  Wolf, and v.~L\"ohneysen}}]{Zocco}
\bibinfo{author}{\bibfnamefont{D.~A.} \bibnamefont{Zocco}},
  \bibinfo{author}{\bibfnamefont{K.}~\bibnamefont{Grube}},
  \bibinfo{author}{\bibfnamefont{F.}~\bibnamefont{Eilers}},
  \bibinfo{author}{\bibfnamefont{T.}~\bibnamefont{Wolf}}, \bibnamefont{and}
  \bibinfo{author}{\bibfnamefont{H.}~\bibnamefont{v.~L\"ohneysen}},
  \bibinfo{journal}{ArXiv e-prints} \textbf{\bibinfo{volume}{1305.5130}}
  (\bibinfo{year}{2013}).

\bibitem[{\citenamefont{Liu et~al.}(2013)\citenamefont{Liu, Tanatar, Kogan,
  Kim, Lograsso, and Prozorov}}]{Liu13}
\bibinfo{author}{\bibfnamefont{Y.}~\bibnamefont{Liu}},
  \bibinfo{author}{\bibfnamefont{M.~A.} \bibnamefont{Tanatar}},
  \bibinfo{author}{\bibfnamefont{V.~G.} \bibnamefont{Kogan}},
  \bibinfo{author}{\bibfnamefont{H.}~\bibnamefont{Kim}},
  \bibinfo{author}{\bibfnamefont{T.~A.} \bibnamefont{Lograsso}},
  \bibnamefont{and} \bibinfo{author}{\bibfnamefont{R.}~\bibnamefont{Prozorov}},
  \bibinfo{journal}{Phys. Rev. B} \textbf{\bibinfo{volume}{87}},
  \bibinfo{pages}{134513} (\bibinfo{year}{2013}).

\bibitem[{\citenamefont{Altarawneh et~al.}(2008)\citenamefont{Altarawneh,
  Collar, Mielke, Ni, Bud'ko, and Canfield}}]{Altarawneh08}
\bibinfo{author}{\bibfnamefont{M.~M.} \bibnamefont{Altarawneh}},
  \bibinfo{author}{\bibfnamefont{K.}~\bibnamefont{Collar}},
  \bibinfo{author}{\bibfnamefont{C.~H.} \bibnamefont{Mielke}},
  \bibinfo{author}{\bibfnamefont{N.}~\bibnamefont{Ni}},
  \bibinfo{author}{\bibfnamefont{S.~L.} \bibnamefont{Bud'ko}},
  \bibnamefont{and} \bibinfo{author}{\bibfnamefont{P.~C.}
  \bibnamefont{Canfield}}, \bibinfo{journal}{Phys. Rev. B}
  \textbf{\bibinfo{volume}{78}}, \bibinfo{pages}{220505}
  (\bibinfo{year}{2008}).

\bibitem[{\citenamefont{Lyard et~al.}(2004)\citenamefont{Lyard, Szab\'o, Klein,
  Marcus, Marcenat, Kim, Kang, Lee, and Lee}}]{Lyard04}
\bibinfo{author}{\bibfnamefont{L.}~\bibnamefont{Lyard}},
  \bibinfo{author}{\bibfnamefont{P.}~\bibnamefont{Szab\'o}},
  \bibinfo{author}{\bibfnamefont{T.}~\bibnamefont{Klein}},
  \bibinfo{author}{\bibfnamefont{J.}~\bibnamefont{Marcus}},
  \bibinfo{author}{\bibfnamefont{C.}~\bibnamefont{Marcenat}},
  \bibinfo{author}{\bibfnamefont{K.~H.} \bibnamefont{Kim}},
  \bibinfo{author}{\bibfnamefont{B.~W.} \bibnamefont{Kang}},
  \bibinfo{author}{\bibfnamefont{H.~S.} \bibnamefont{Lee}}, \bibnamefont{and}
  \bibinfo{author}{\bibfnamefont{S.~I.} \bibnamefont{Lee}},
  \bibinfo{journal}{Phys. Rev. Lett.} \textbf{\bibinfo{volume}{92}},
  \bibinfo{pages}{057001} (\bibinfo{year}{2004}).

\bibitem[{\citenamefont{Helfand and Werthamer}(1964)}]{WHH1}
\bibinfo{author}{\bibfnamefont{E.}~\bibnamefont{Helfand}} \bibnamefont{and}
  \bibinfo{author}{\bibfnamefont{N.~R.} \bibnamefont{Werthamer}},
  \bibinfo{journal}{Phys. Rev. Lett.} \textbf{\bibinfo{volume}{13}},
  \bibinfo{pages}{686} (\bibinfo{year}{1964}).

\bibitem[{\citenamefont{Helfand and Werthamer}(1966)}]{WHH2}
\bibinfo{author}{\bibfnamefont{E.}~\bibnamefont{Helfand}} \bibnamefont{and}
  \bibinfo{author}{\bibfnamefont{N.~R.} \bibnamefont{Werthamer}},
  \bibinfo{journal}{Phys. Rev.} \textbf{\bibinfo{volume}{147}},
  \bibinfo{pages}{288} (\bibinfo{year}{1966}).

\bibitem[{\citenamefont{Brison et~al.}(1995)\citenamefont{Brison, Keller,
  Verniere, Lejay, Schmidt, Buzdin, Flouquet, Julian, and
  Lonzarich}}]{Brison95}
\bibinfo{author}{\bibfnamefont{J.~P.} \bibnamefont{Brison}},
  \bibinfo{author}{\bibfnamefont{N.}~\bibnamefont{Keller}},
  \bibinfo{author}{\bibfnamefont{A.}~\bibnamefont{Verniere}},
  \bibinfo{author}{\bibfnamefont{P.}~\bibnamefont{Lejay}},
  \bibinfo{author}{\bibfnamefont{L.}~\bibnamefont{Schmidt}},
  \bibinfo{author}{\bibfnamefont{A.}~\bibnamefont{Buzdin}},
  \bibinfo{author}{\bibfnamefont{J.}~\bibnamefont{Flouquet}},
  \bibinfo{author}{\bibfnamefont{S.~R.} \bibnamefont{Julian}},
  \bibnamefont{and} \bibinfo{author}{\bibfnamefont{G.~G.}
  \bibnamefont{Lonzarich}}, \bibinfo{journal}{Physica C}
  \textbf{\bibinfo{volume}{250}}, \bibinfo{pages}{128} (\bibinfo{year}{1995}).

\bibitem[{\citenamefont{Saint-James et~al.}(1969)\citenamefont{Saint-James,
  Thomas, and Sarma}}]{SaintJamesBook}
\bibinfo{author}{\bibfnamefont{D.}~\bibnamefont{Saint-James}},
  \bibinfo{author}{\bibfnamefont{E.~J.} \bibnamefont{Thomas}},
  \bibnamefont{and} \bibinfo{author}{\bibfnamefont{G.}~\bibnamefont{Sarma}},
  \emph{\bibinfo{title}{Type II Superconductivity}}
  (\bibinfo{publisher}{Pergamon Press}, \bibinfo{year}{1969}).

\bibitem[{\citenamefont{Orlando et~al.}(1979)\citenamefont{Orlando,
  E.~J.~McNiff, Foner, and Beasley}}]{Orlando79}
\bibinfo{author}{\bibfnamefont{T.~P.} \bibnamefont{Orlando}},
  \bibinfo{author}{\bibfnamefont{J.}~\bibnamefont{E.~J.~McNiff}},
  \bibinfo{author}{\bibfnamefont{S.}~\bibnamefont{Foner}}, \bibnamefont{and}
  \bibinfo{author}{\bibfnamefont{M.~R.} \bibnamefont{Beasley}},
  \bibinfo{journal}{Phys. Rev. B} \textbf{\bibinfo{volume}{19}},
  \bibinfo{pages}{4545} (\bibinfo{year}{1979}).

\bibitem[{\citenamefont{Yang and Sondhi}(1998)}]{Yang98}
\bibinfo{author}{\bibfnamefont{K.}~\bibnamefont{Yang}} \bibnamefont{and}
  \bibinfo{author}{\bibfnamefont{S.~L.} \bibnamefont{Sondhi}},
  \bibinfo{journal}{Phys. Rev. B} \textbf{\bibinfo{volume}{57}},
  \bibinfo{pages}{8566} (\bibinfo{year}{1998}).

\bibitem[{\citenamefont{Barzykin}(2009)}]{Barzykin09}
\bibinfo{author}{\bibfnamefont{V.}~\bibnamefont{Barzykin}},
  \bibinfo{journal}{Phys. Rev. B} \textbf{\bibinfo{volume}{79}},
  \bibinfo{pages}{134517} (\bibinfo{year}{2009}).

\bibitem[{\citenamefont{Won et~al.}(1999)\citenamefont{Won, Jang, and
  Maki}}]{Won99}
\bibinfo{author}{\bibfnamefont{H.}~\bibnamefont{Won}},
  \bibinfo{author}{\bibfnamefont{H.}~\bibnamefont{Jang}}, \bibnamefont{and}
  \bibinfo{author}{\bibfnamefont{K.}~\bibnamefont{Maki}},
  \bibinfo{journal}{ArXiv e-prints} \textbf{\bibinfo{volume}{9901252}}
  (\bibinfo{year}{1999}).

\bibitem[{\citenamefont{Barzykin and Gor'kov}(2007)}]{Barzykin07}
\bibinfo{author}{\bibfnamefont{V.}~\bibnamefont{Barzykin}} \bibnamefont{and}
  \bibinfo{author}{\bibfnamefont{L.~P.} \bibnamefont{Gor'kov}},
  \bibinfo{journal}{Phys. Rev. Lett.} \textbf{\bibinfo{volume}{98}},
  \bibinfo{pages}{087004} (\bibinfo{year}{2007}).

\bibitem[{\citenamefont{Machida and Ichioka}(2008)}]{Machida08}
\bibinfo{author}{\bibfnamefont{K.}~\bibnamefont{Machida}} \bibnamefont{and}
  \bibinfo{author}{\bibfnamefont{M.}~\bibnamefont{Ichioka}},
  \bibinfo{journal}{Phys. Rev. B} \textbf{\bibinfo{volume}{77}},
  \bibinfo{pages}{184515} (\bibinfo{year}{2008}).

\bibitem[{\citenamefont{Gruenberg and Gunther}(1966)}]{Gruenberg66}
\bibinfo{author}{\bibfnamefont{L.~W.} \bibnamefont{Gruenberg}}
  \bibnamefont{and} \bibinfo{author}{\bibfnamefont{L.}~\bibnamefont{Gunther}},
  \bibinfo{journal}{Phys. Rev. Lett.} \textbf{\bibinfo{volume}{16}},
  \bibinfo{pages}{996} (\bibinfo{year}{1966}).

\bibitem[{\citenamefont{Heid}(2013)}]{Heid13}
\bibinfo{author}{\bibfnamefont{R.}~\bibnamefont{Heid}},
  \bibinfo{journal}{unpublished}  (\bibinfo{year}{2013}).

\bibitem[{\citenamefont{Flouquet et~al.}(1991)\citenamefont{Flouquet, Brison,
  Hasselbach, Taillefer, Behnia, Jaccard, and de~Visser}}]{Flouquet91}
\bibinfo{author}{\bibfnamefont{J.}~\bibnamefont{Flouquet}},
  \bibinfo{author}{\bibfnamefont{J.}~\bibnamefont{Brison}},
  \bibinfo{author}{\bibfnamefont{K.}~\bibnamefont{Hasselbach}},
  \bibinfo{author}{\bibfnamefont{L.}~\bibnamefont{Taillefer}},
  \bibinfo{author}{\bibfnamefont{K.}~\bibnamefont{Behnia}},
  \bibinfo{author}{\bibfnamefont{D.}~\bibnamefont{Jaccard}}, \bibnamefont{and}
  \bibinfo{author}{\bibfnamefont{A.}~\bibnamefont{de~Visser}},
  \bibinfo{journal}{Physica C} \textbf{\bibinfo{volume}{185}},
  \bibinfo{pages}{372} (\bibinfo{year}{1991}).

\end{thebibliography}

\end{document}